\documentclass[letterpaper,twocolumn,prl,showkeys,nofootinbib,longbibliography,reprint,superscriptaddress]{revtex4-1}
\usepackage[T1]{fontenc}
\usepackage[utf8]{inputenc}
\usepackage{color}
\usepackage{mathtools}
\usepackage{amsmath}
\usepackage{amssymb}
\usepackage{graphicx}
\usepackage[unicode=true,bookmarks=false,colorlinks=true]{hyperref}
\hypersetup{linkcolor=Blue, citecolor=Blue, urlcolor=Blue}
\usepackage{breakurl}

\makeatletter

\usepackage{microtype}
\usepackage{stix}
\usepackage[dvipsnames]{xcolor}
\usepackage{siunitx}
\makeatother

\begin{document}

\title{Ultrastrong parametric coupling between a superconducting cavity
and a mechanical resonator}

\author{G. A. Peterson}

\affiliation{National Institute of Standards and Technology, 325 Broadway, Boulder,
CO 80305, USA}

\affiliation{Department of Physics, University of Colorado, Boulder, CO 80309,
USA}

\author{S. Kotler}

\affiliation{National Institute of Standards and Technology, 325 Broadway, Boulder,
CO 80305, USA}

\affiliation{Department of Physics, University of Colorado, Boulder, CO 80309,
USA}

\author{F. Lecocq}

\affiliation{National Institute of Standards and Technology, 325 Broadway, Boulder,
CO 80305, USA}

\affiliation{Department of Physics, University of Colorado, Boulder, CO 80309,
USA}

\author{K. Cicak}

\affiliation{National Institute of Standards and Technology, 325 Broadway, Boulder,
CO 80305, USA}

\author{X. Y. Jin}

\affiliation{National Institute of Standards and Technology, 325 Broadway, Boulder,
CO 80305, USA}

\affiliation{Department of Physics, University of Colorado, Boulder, CO 80309,
USA}

\author{R. W. Simmonds}

\affiliation{National Institute of Standards and Technology, 325 Broadway, Boulder,
CO 80305, USA}

\author{J. Aumentado}

\affiliation{National Institute of Standards and Technology, 325 Broadway, Boulder,
CO 80305, USA}

\author{J. D. Teufel}

\affiliation{National Institute of Standards and Technology, 325 Broadway, Boulder,
CO 80305, USA}
\begin{abstract}

We present a new optomechanical device where the motion of a micromechanical
membrane couples to a microwave resonance of a three-dimensional superconducting
cavity. With this architecture, we realize ultrastrong parametric
coupling, where the coupling rate not only exceeds the dissipation
rates in the system but also rivals the mechanical frequency itself.
In this regime, the optomechanical interaction induces a frequency
splitting between the hybridized normal modes that reaches 88\% of
the bare mechanical frequency, limited by the fundamental parametric
instability. The coupling also exceeds the mechanical thermal decoherence
rate, enabling new applications in ultrafast quantum state transfer
and entanglement generation.

\end{abstract}

\date{\today}

\keywords{cavity optomechanics, superconducting circuits, ultrastrong coupling}

\maketitle

The physics of coupled oscillators is used to understand a wide array
of natural and man-made phenomena, from the microscopic vibrations
of atoms and molecules to the interplay of planets and their moons.
Coupling strengths can be grouped into different regimes that entail
qualitatively different behavior. In particular, the regimes of weak
and strong coupling are distinguished by whether or not the coupling
between two oscillators exceeds their damping rates. In the strong
coupling regime, the eigenfrequencies of the combined system split
into normal modes where the energy swaps back and forth between the
individual oscillators. For low-loss systems, this energy exchange
can be fast compared to the lifetimes of the individual modes while
still remaining slow compared to their periods. The strong coupling
regime has become an essential tool in engineered quantum systems
because it can allow the subsystems to exchange their quantum information
before it is lost due to decoherence \cite{MabuchiDoherty}. As quantum
devices continue to be engineered with increased coupling rates, a
new regime known as ultrastrong coupling has become relevant. This
regime occurs once the coupling rate becomes so large as to rival
a bare resonance frequency, resulting in new quantum effects including
multimode entanglement and virtually excited ground states \cite{NoriKockum_NatRev_2019,SolanoForn_RMP_2019}.
Reaching ultrastrong coupling and studying its effect in quantum devices
remains an active experimental challenge.

Cavity optomechanics is an area of engineered quantum systems in which
a mechanical resonator and an electromagnetic mode form a coupled-oscillator
system \cite{AspelmeyerKippenbergMarquardt2014}. Although the intrinsic
coupling rate between single photons and phonons is typically small,
cavity optomechanical systems allow an enhancement of the coupling
proportional to the amplitude of a coherent cavity drive. This parametric
enhancement has allowed demonstrations of strong coupling both at
ambient temperatures \cite{Aspelmeyer_Nature_2009,Vanner_Optica_2019}
and in cryogenic, quantum-coherent regimes \cite{TeufelLi_Nature_2011,KippenbergVerhagen_Nature_2012,NakamuraNoguchi_NJP_2016,Rakich_arxiv_2018}.
In practice, as the intensity of the coherent drive becomes large,
it can induce other undesirable effects including heating and cavity
nonlinearity, limiting the final parametric coupling. Operating deeply
within the quantum-coherent ultrastrong coupling regime therefore
requires a dramatic increase in either the single-photon optomechanical
coupling rate or the cavity's power handling capability. Specifically,
in the microwave domain, one prominent optomechanical platform is
a lumped element superconducting circuit formed from a mechanically
compliant vacuum-gap capacitor shunted by a thin-film inductor. While
this architecture has enabled strong coupling \cite{TeufelLi_Nature_2011},
ground state cooling \cite{TeufelDonnerLiEtAl2011}, and entanglement
\cite{LehnertPalomaki2013}, the coupling rate has remained well below
the onset of ultrastrong coupling effects, limited by unwanted nonlinearity
of the superconducting inductor \cite{Zmuidzinas2012}.

\begin{figure}
\begin{centering}
\includegraphics{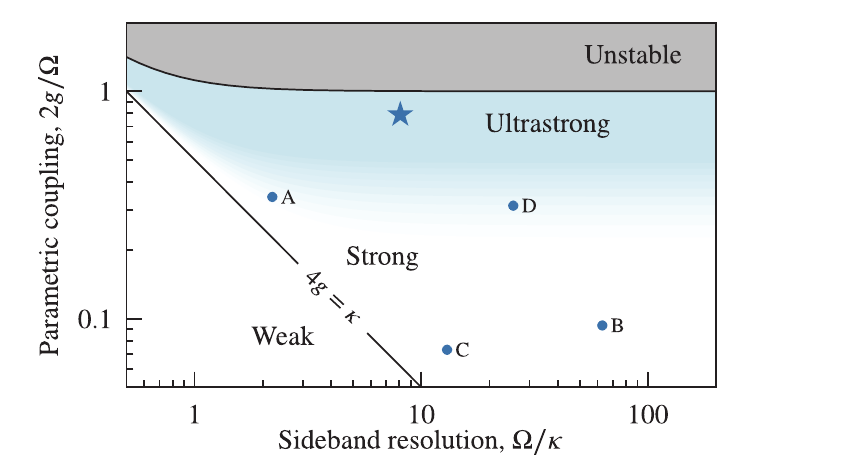}
\par\end{centering}
\caption{Parameter space diagram showing four regimes of parametric optomechanical
coupling $g$ as a function of cavity dissipation $\kappa$ and mechanical
frequency $\Omega$. For a parametric drive at $\Delta=-\Omega$,
strong coupling coincides with the normal-mode splitting condition
$\kappa<4g$. Ultrastrong coupling arises when $g$ further increases
to a significant fraction of $\Omega$ until reaching the limit for
stability at $2g=\sqrt{\Omega^{2}+\kappa^{2}/4}$. Labeled points
denote previous optomechanical experiments in the strong coupling
regime: (A) optical Fabry–Pérot cavity \cite{Aspelmeyer_Nature_2009},
(B) lumped element microwave circuit \cite{TeufelLi_Nature_2011},
(C) toroidal optical microcavity \cite{KippenbergVerhagen_Nature_2012},
and (D) microwave loop-gap cavity \cite{NakamuraNoguchi_NJP_2016}.
The star indicates the highest coupling rate achieved in this work.
\label{fig:phase_diagram}}
\end{figure}

In this Letter, we introduce a new optomechanical architecture that
mitigates the nonideality of previous designs, allowing us to reach
ultrastrong coupling and approach the fundamental stability limit
of the pure optomechanical interaction \cite{Nunnenkamp_PRA_2013}.
Our device consists of a microfabricated vacuum-gap capacitor embedded
in a three-dimensional superconducting microwave cavity, analogous
to recent work in the field of circuit quantum electrodynamics \cite{SchoelkopfPaik_PRL}
and similar to other optomechanical demonstrations \cite{BlairTobar1993,SteeleYuan_NatComm_2015,NakamuraNoguchi_NJP_2016,SinghGunupdi_PRApp_2019}.
Our device takes advantage of the superior power handling of bulk
cavity resonators compared to thin-film inductors. In general, the
drawback of using a cavity resonator is a larger parasitic capacitance
that would dilute the optomechanical coupling. Crucially, through
careful microwave design and simulation, we maintain the relatively
large single-photon coupling of lumped element vacuum-gap circuits
\cite{TeufelLi_Nature_2011}. As a result, we achieve ultrastrong
parametric coupling by applying microwave drives with one hundred
times larger power, ultimately limited by the instability inherent
in the optomechanical Hamiltonian. Due to the low temperature operation,
the quantum decoherence rates are kept sufficiently small to enable
new regimes of entanglement \cite{Nunnenkamp_PRA_2013,HammererPRA2015},
nonlinear quantum optomechanics \cite{ClerkLemonde_PRL_2013}, and
ultrafast quantum state transfer \cite{LehnertPalomaki_Nature_2013,SimmondsLecocq_natphys_2015,LehnertReed_NatPhys_2017}.

\begin{figure}
\centering{}\includegraphics{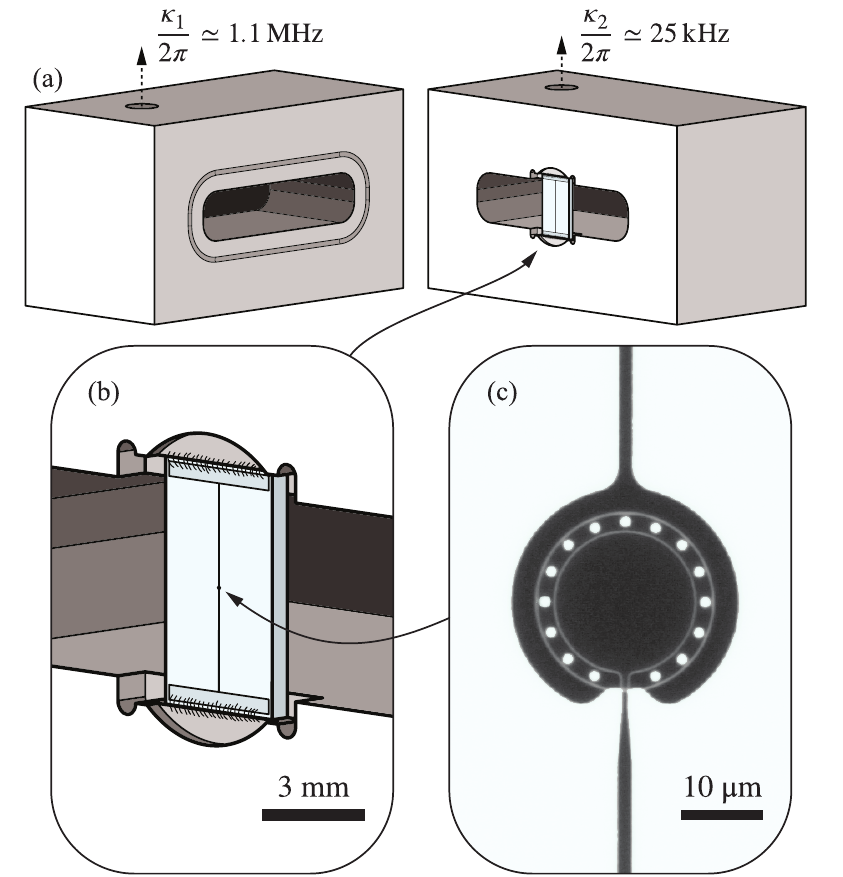} \caption{Device schematic. (a) A cavity, milled from two aluminum blocks, supports
a microwave resonance near $\SI{12}{\giga\hertz}$. Microwave signals
couple in and out of the cavity through two asymmetrically coupled
ports. (b) An aluminum mechanically compliant capacitor patterned
on a sapphire substrate is galvanically connected to the cavity walls
through superconducting wirebonds, loading the fundamental cavity
resonance frequency down to $\SI{6.5}{\giga\hertz}$. (c) A micrograph
image shows the capacitor, which has a fundamental mechanical resonance
at $\SI{9.7}{\mega\hertz}$ and a vacuum gap of approximately $\SI{30}{\nano\meter}$.
\label{fig:diagram}}
\end{figure}

In a generic cavity optomechanical system, the natural frequency $\omega_{c}$
of an electromagnetic resonance depends on the position $\hat{x}$
of a mechanical harmonic oscillator \cite{AspelmeyerKippenbergMarquardt2014}.
The interaction Hamiltonian is $\hat{H}_{\text{int}}=\hslash g_{0}\hat{n}\hat{x}/x_{zp}$,
where $g_{0}$ is the single-photon coupling rate, $\hat{n}$ is the
number operator for microwave photons, and $\hslash$ is the reduced
Planck constant. $x_{zp}=\sqrt{\hslash/2m\Omega}$ is the mechanical
zero-point fluctuation amplitude, where $m$ and $\Omega$ are the
effective mass and the resonance frequency of the mechanical mode,
respectively. Dissipation in the system is characterized by damping
rates $\kappa$ for the electrical mode and $\Gamma$ for the mechanical
mode, with $\Gamma\ll\kappa$. Even if $g_{0}$ is small, as is the
case in most optomechanical systems, the coupling can be parametrically
enhanced by driving the electrical mode to a coherent state of mean
photon number $n_{d}$ at frequency $\omega_{d}$. Defining $\hat{a}$
as the annihilation operator for fluctuations around the driven state,
we can approximate the interaction Hamiltonian as $\hat{H}_{\text{int}}=\hslash g(\hat{a}+\hat{a}^{\dagger})\hat{x}/x_{zp}$,
where $g=g_{0}\sqrt{n_{d}}$ is the parametrically enhanced coupling
rate, leading to a set of linear coupled equations of motion for $\hat{a}(t)$
and $\hat{x}(t)$. Just as two passively coupled oscillators interact
most strongly when resonant with each other, for parametric coupling
an effective resonance condition is reached when driving the system
at the difference frequency ($\Delta\equiv\omega_{d}-\omega_{c}=-\Omega$)
in the resolved sideband regime ($\kappa\ll\Omega$). These conditions
optimize the coherent exchange of energy between the mechanical and
electromagnetic modes.

As the driven coupling $g$ is increased from its small single-photon
value, we encounter several distinct regimes of coupling. When the
cooperativity $C=4g^{2}/\kappa\Gamma$ reaches $1$, the optical damping
of the mechanical mode begins to dominate over its intrinsic dissipation.
As $g$ increases further, the effective mechanical linewidth increases
until it reaches the cavity dissipation rate when $g=\kappa/4$. Above
this threshold, the system enters the strong-coupling regime where
the cavity and mechanical mode hybridize, with the mechanical resonance
frequency splitting into two solutions, which for $g\ll\Omega$ are
given by 
\begin{equation}
\Omega_{\pm}\simeq\Omega\pm\sqrt{g^{2}-\kappa^{2}/16}.\label{eq:SC_approx}
\end{equation}
This leads to the splitting frequency $\Omega_{s}=\Omega_{+}-\Omega_{-}\simeq2g$
when the coupling overwhelms the cavity dissipation. In this regime
of strong coupling, the two physical resonators exchange energy and
information at a rate $\Omega_{s}$, faster than any dissipation in
the system.

As the splitting frequency $\Omega_{s}$ approaches the bare mechanical
frequency $\Omega$, however, counter-rotating terms of order $g/\Omega$
cannot be ignored, requiring the use of the exact eigenfrequency spectrum
\cite{Supplement}
\begin{equation}
\Omega_{\pm}=\mathrm{Re}\sqrt{\Omega^{2}-\frac{\kappa^{2}}{16}\pm2\Omega\sqrt{g^{2}-\frac{\kappa^{2}}{16}}}.\label{eq:mech_normal_freqs}
\end{equation}
The discrepancy between \eqref{eq:SC_approx} and \eqref{eq:mech_normal_freqs}
is a measurable metric to distinguish strong and ultrastrong parametric
coupling. In the resolved sideband ultrastrong coupling regime ($\kappa\ll2g<\Omega$),
Eq.~\eqref{eq:mech_normal_freqs} becomes $\Omega_{\pm}\simeq\Omega\sqrt{1\pm2g/\Omega}$,
showing that the splitting exceeds $2g$ until the system becomes
parametrically unstable when $\Omega_{-}=0$ at $g=\Omega/2$, corresponding
to a splitting frequency $\Omega_{s}=\sqrt{2}\Omega$. 

The regimes discussed above are shown in Fig.~\ref{fig:phase_diagram}
as a function of parametric coupling and sideband resolution $\Omega/\kappa$.
The shaded region of ultrastrong coupling corresponds to $\Omega/5<\Omega_{s}$,
roughly where terms of order $g/\Omega$ become relevant while still
satisfying the condition for strong coupling. As $\Omega_{s}$ characterizes
the rate at which the two physical resonators exchange energy, the
instability sets a fundamental limit for both optomechanical coupling
as well as coherent exchange of information between the resonators
in the steady state. Reaching ultrastrong coupling therefore allows
the exploration of the fundamental limitations of coupling in optomechanical
systems.

For quantum applications, the coupling rate should be compared not
only to dissipation but also to the decoherence rates in the system.
In particular, the mechanical thermal decoherence rate $n_{\text{th}}\Gamma$
can be much larger than the intrinsic dissipation $\Gamma$, where
$n_{\text{th}}$ is the mechanical occupancy when in equilibrium with
its thermal environment. Ideally, the quantum cooperativity $C_{q}=4g^{2}/\kappa n_{\text{th}}\Gamma$
exceeds 1 before the onset of strong coupling, ensuring that the hybridized
system is quantum coherent. In the following, we present an optomechanical
device that achieves the hierarchy of rates desired for quantum coherent
ultrastrong coupling: $n_{\text{th}}\Gamma\ll\kappa/2\ll2g\lesssim\Omega\ll\omega_{c}$.

Our device is shown in Fig.~\ref{fig:diagram}. A microwave cavity
resonator with inner dimensions $\SI{19}{\milli\meter}\times\SI{4}{\milli\meter}\times\SI{17}{\milli\meter}$
milled from bulk aluminum defines the electrical resonance of the
system. We focus on the fundamental $\mathrm{TE}_{101}$ microwave
mode, whose electric field is maximal at the center of the cavity,
where we place a sapphire chip containing a microfabricated $\SI{20}{\micro\meter}$-diameter
aluminum vacuum-gap capacitor \cite{SimmondsLecocq_natphys_2015}.
The suspended top plate of the capacitor forms the mechanical resonator
of the system. Reducing the parasitic capacitance of the cavity and
thereby maximizing the optomechanical coupling rate requires a galvanic
connection between the microfabricated capacitor and the cavity walls.
To achieve this, we use aluminum bond wires to connect the cavity
faces to lithographically patterned pads, which then connect to the
capacitor through thin-film aluminum wires. 

The vacuum-gap capacitor and sapphire substrate load the cavity resonance,
pulling its frequency from around $\SI{12}{\giga\hertz}$ down to
$\omega_{c}/2\pi\approx\SI{6.506}{\giga\hertz}$. Two cavity ports
with adjustable coupling pins allow signals to couple in and out to
coaxial cables. We adjust the length of the pins at room temperature
so the two ports contribute asymmetrically to the total dissipation
$\kappa=\kappa_{1}+\kappa_{2}+\kappa_{i}$, where $\kappa_{1}/2\pi\approx\SI{1.1}{\mega\hertz}$
and $\kappa_{2}/2\pi\approx\SI{25}{\kilo\hertz}$ are the port coupling
rates. The internal dissipation $\kappa_{i}$ of the cavity mode ranges
from $\SI{\sim30}{\kilo\hertz}$ to $\SI{\sim140}{\kilo\hertz}$ depending
on the circulating power \cite{Martinis_PRL_2005}. The cavity is
therefore overcoupled with a total dissipation rate $\kappa/2\pi\approx\SI{1.2}{\mega\hertz}$.
We place the device in a cryostat with a base temperature of $\SI{16}{\milli\kelvin}$
and probe and monitor the system with microwave signals applied near
the cavity resonance frequency. Our setup allows us to measure all
four elements of the scattering matrix for a broad range of parametric
coupling parameters. With a weak cavity drive, we characterize the
fundamental vibrational mode of the capacitor plate by its resonance
frequency $\Omega/2\pi=\SI{9.696}{\mega\hertz}$ and its intrinsic
damping rate $\Gamma/2\pi=\SI{31\pm1}{\hertz}$, measured spectroscopically,
where the uncertainty represents the standard error of the mean. By
varying the cryostat temperature and measuring mechanical thermal
noise, we determine the single-photon optomechanical coupling rate
$g_{0}/2\pi=\SI{167\pm2}{\hertz}$ \cite{TeufelDonnerLiEtAl2011}.
We also find that at base temperature, the mechanical mode equilibrates
to $\SI{35\pm3}{\milli\kelvin}$, corresponding to a thermal phonon
occupancy $n_{\text{th}}=\SI{76\pm6}{}$.

\begin{figure}
\centering{}\includegraphics{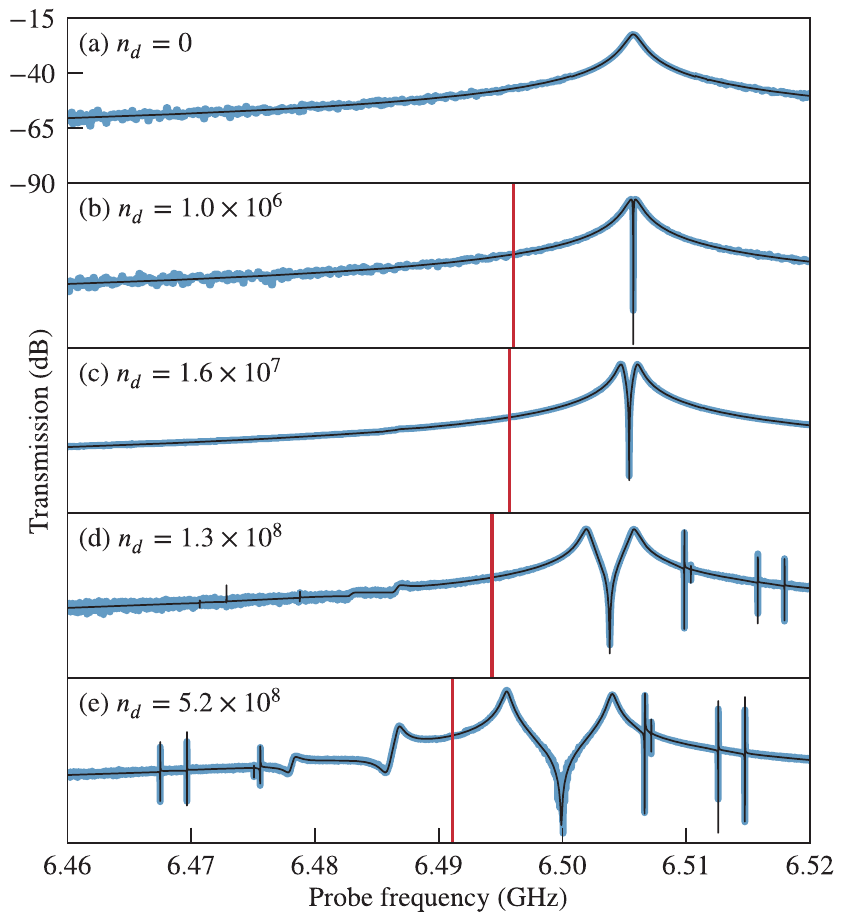}\caption{Measured and calibrated cavity transmission from port one to port
two for varied drive strengths $n_{d}$ from weak through strong coupling
and up to ultrastrong coupling. The data (blue) is fit to theory (black)
containing the first five mechanical modes. The vertical red line
indicates the frequency of the applied microwave drive, which is adjusted
with power to maintain $\Delta=-\Omega$. The structure below the
drive frequency at the highest powers directly shows the importance
of counter-rotating terms in the ultrastrong coupling regime. \label{fig:Fig2}}
\end{figure}

To probe the response of the coupled system, we measure the vector
transmission of the microwave field through the cavity in the presence
of a parametric drive \cite{KippenbergWeis_Science_2010,TeufelLi_Nature_2011,PainterSafaviNaeini_Nature_2011}.
In Fig.~\ref{fig:Fig2}, we plot the magnitude of the transmission
for a range of drive powers applied below the cavity resonance, with
the drive frequency indicated as a red vertical line. We coherently
cancel the drive after it leaves the cavity, allowing us to measure
the transmission of a weak probe without saturating our microwave
measurement \cite{Supplement}. We adjust the drive frequency as we
increase drive power to maintain the condition $\Delta=-\Omega$ in
the presence of two dominant nonlinear effects. Namely, we measure
the pure optomechanical Kerr shift, $-2g_{0}^{2}/\Omega=\SI{-5.8\pm0.1}{\milli\hertz\per photon}$,
and we attribute the remaining shift to the residual nonlinear kinetic
inductance of the superconducting film, approximately $\SI{-4}{\milli\hertz\per photon}$
at our highest powers \cite{Zmuidzinas2012}.

At low drive power (Fig.~\ref{fig:Fig2}a), we measure the bare cavity
resonance, a single Lorentzian with linewidth $\kappa$. As the drive
power increases (Fig.~\ref{fig:Fig2}b), the optomechanical interaction
appears as interference in the cavity response with a characteristic
bandwidth of the damped mechanical linewidth $\sim4g^{2}/\kappa$.
At large enough power (Fig.~\ref{fig:Fig2}c), the damped mechanical
width reaches $\kappa/2$, after which the response splits into normal
modes, marking the strong coupling regime. Additionally, we begin
to resolve the next four vibrational modes of the membrane (Fig.~\ref{fig:Fig2}d),
which couple weakly to the cavity mode. At the highest power (Fig.~\ref{fig:Fig2}e),
the response acquires several features indicative of ultrastrong coupling.
In addition to the splitting of the fundamental resonance becoming
of order $\Omega$, the counter-rotating dynamics below the drive
frequency become significant and easily observable. Lastly, the transmission
at the drive frequency begins to increase, signifying a nonlinear
relationship between input power and driven photon number in the cavity.

\begin{figure}
\begin{centering}
\includegraphics{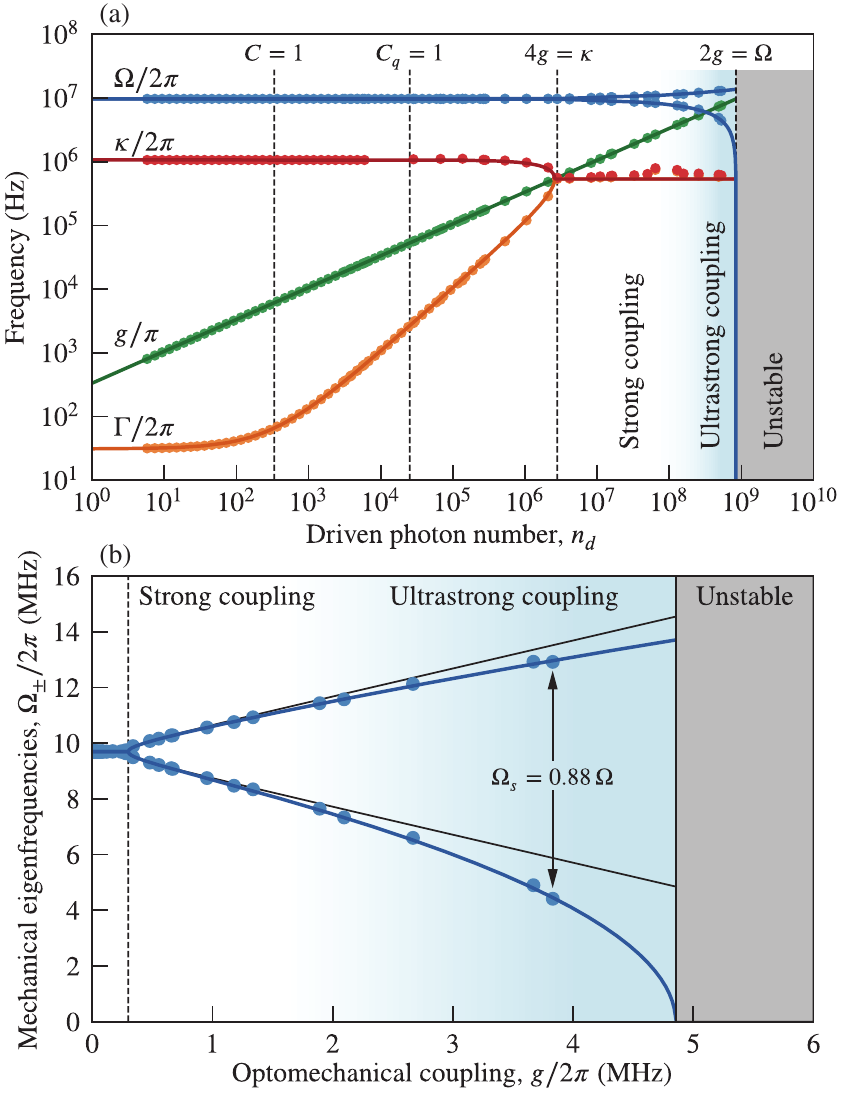}
\par\end{centering}
\begin{centering}
\caption{(a) Measured system rates as a function of drive strength. In the
weak coupling regime, the optical damping first exceeds the mechanical
dissipation rate ($C=1$) and then exceeds the thermal decoherence
rate ($C_{q}=1$). Once $2g$ reaches $\kappa/2$, the eigenmodes
split, indicating the onset of strong coupling. Eventually, ultrastrong
coupling corrections become important before the system approaches
a parametric instability at $2g=\Omega$. (b) Mechanical eigenfrequencies
as a function of the optomechanical coupling rate. In the ultrastrong
coupling regime, the splitting $\Omega_{s}=\Omega_{+}-\Omega_{-}$
exceeds the linear strong-coupling approximation $\Omega_{s}\simeq2g$
(black), reaching a maximal value $\Omega_{s}\approx0.88\,\Omega$.
\label{fig:Fig3}}
\par\end{centering}
\end{figure}

We fit the data in the complex plane to multimode optomechanical theory
\cite{Supplement}, shown as a black line plotted over the data. The
quantitative agreement between data and theory allows us to extract
all the relevant system parameters as a function of drive power. These
parameters are shown in Fig.~\ref{fig:Fig3}a from very weak parametric
coupling to the ultrastrong coupling regime. For very few drive photons
(in this case $n_{d}\lesssim10^{2}$), the bare mode properties are
measured. As the drive power increases within the regime of weak coupling
($n_{d}\lesssim10^{6}$), the mechanical mode is damped and cooled,
passing through $C=1$ and entering the quantum-enabled regime $C_{q}>1$,
where the occupancy is reduced below one quantum. Normal-mode splitting
occurs at $n_{d}\approx3\times10^{6}$ where $4g=\kappa$, marking
the beginning of strong coupling. Above $n_{d}\approx2\times10^{8}$,
the splitting starts to deviate from $2g$ as the system enters the
ultrastrong coupling regime. The threshold for parametric oscillation
occurs at $n_{d}\approx8.4\times10^{8}$, above which no steady-state
solution exists. We measure well into the regime where $\Omega_{-}<2g<\Omega_{s}$;
that is, the energy swapping rate exceeds $2g$ as well as the lower
eigenfrequency itself.

In Fig.~\ref{fig:Fig3}b, the measured mechanical eigenfrequencies
(blue circles) are plotted versus the optomechanical coupling rate.
The black line shows the strong-coupling approximation \eqref{eq:SC_approx},
while the full solution \eqref{eq:mech_normal_freqs} is shown as
a blue line. The discrepancy between the two is a clear indication
of ultrastrong coupling effects. At the highest power, the splitting
between the two modes exceeds the frequency of the lower mode, reaching
88\% of the bare mechanical frequency.

\begin{figure}
\begin{centering}
\includegraphics{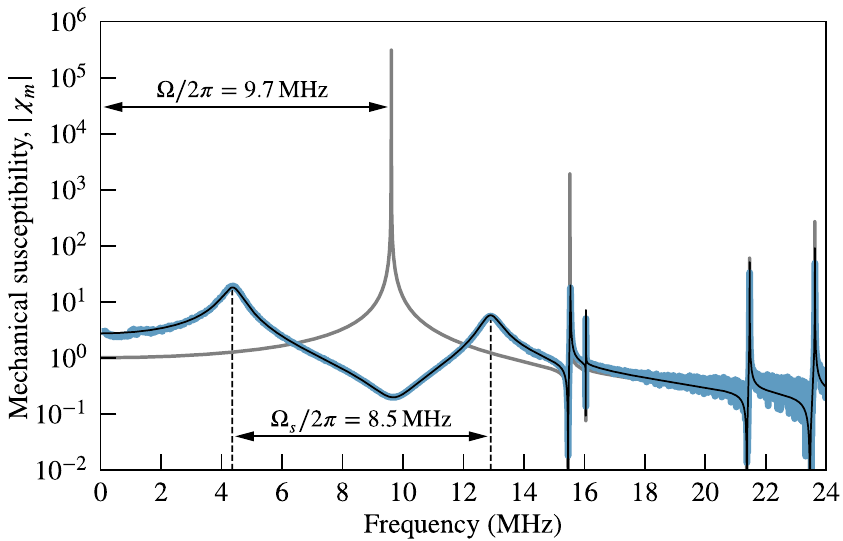}
\par\end{centering}
\caption{Mechanical susceptibility inferred from microwave measurements at
our highest drive power (blue) with fit line (black) compared with
the bare susceptibility (gray). In agreement with the full theory,
the susceptibility in the ultrastrong coupling regime shows asymmetry
at the two eigenfrequencies as well as a marked increase in its value
at zero frequency. Here, the splitting is $\SI{8.5}{\mega\hertz}$
with a corresponding cooperativity of $C\approx1.6\times10^{6}$.
\label{fig:Mechanical-susceptibility}}
\end{figure}

Our measurements of the driven cavity response allow us to reconstruct
the mechanical susceptibility, defined as the ratio of induced motion
to external force $\chi_{m}(\omega)=m\Omega^{2}x(\omega)/F(\omega)$,
normalized by the intrinsic mechanical spring constant $m\Omega^{2}$.
This is shown in Fig.~\ref{fig:Mechanical-susceptibility}. The gray
line corresponds to the estimated susceptibility at zero coupling,
which includes the bare resonance of the fundamental mode as well
as the first four higher-order vibrational modes. The peak height
on resonance is equal to the quality factor of the fundamental mode,
$Q_{m}=\Omega/\Gamma\approx3\times10^{5}$, while the heights at the
other resonances are reduced due to their lower vacuum coupling rates.
The data plotted in blue (with a fit to theory in black) is the susceptibility
inferred from our highest drive power in Fig.~\ref{fig:Fig2}. The
original peak of linewidth $\Gamma$ is split into two broad peaks
with approximate widths $\kappa/2$, each representing a normal mode
of the joint cavity-mechanics system. At this large coupling rate,
the eigenfrequencies are no longer split symmetrically about the bare
mechanical frequency, and the peak heights are unequal. As the lower
eigenfrequency decreases, we also see an increase in the value of
susceptibility at zero frequency, indicating an increased response
to static forces. All of these effects agree with the full optomechanical
theory that includes ultrastrong-coupling corrections. At our highest
cooperativity of $C\approx1.6\times10^{6}$, we achieve a maximal
splitting of $\Omega_{s}/2\pi\approx\SI{8.5}{\mega\hertz}$. Here,
the state of the mechanical mode is exchanged with the cavity mode
in a characteristic time $\pi/\Omega_{s}\approx\SI{60}{\nano\second}$,
faster than the mechanical oscillation period $2\pi/\Omega\approx\SI{100}{\nano\second}$.

In conclusion, we have introduced a novel microwave optomechanical
circuit architecture and experimentally characterized the eigenmodes
of the system in the quantum regime. As the parametric coupling is
increased \emph{in situ}, we find quantitative agreement with the
theoretical predictions from weak to ultrastrong coupling. Looking
forward, ultrastrong coupling of bosonic modes is predicted to have
interesting ground state properties \cite{Carusotto_PRB_2005,CarusottoCiuti_PRA,MilmanFedortchenko_PRA,Huard_PRL_2018}.
Quantum correlations induced by the counter-rotating terms in the
ultrastrong coupling regime induce squeezing and entanglement between
the mechanical and cavity modes \cite{HammererPRA2015}. Studying
the noise properties in this regime would also allow us to observe
these effects for the first time, as well as assess how these correlations
could be exploited for quantum-enhanced sensing of forces, displacements,
or gravitational waves \cite{LamSchabel_NatComm_2010,RegalKampel_PRX_2017,SchliesserMason_NatPhys_2019}.
Furthermore, while the fundamental instability precludes traditional
steady-state measurements when the swapping rate exceeds the mechanical
frequency, pulsed measurements would allow the system to be characterized
beyond this limit into the deep strong coupling regime \cite{NoriKockum_NatRev_2019,SolanoForn_RMP_2019}.
These ultrafast pulsed measurements would allow new regimes of quantum
state transfer \cite{LehnertPalomaki_Nature_2013,LehnertReed_NatPhys_2017},
entanglement \cite{LehnertPalomaki2013}, topologically-protected
operations \cite{HarrisXu_Nature_2016}, and projective measurements
\cite{AspelmeyerVanner_PNAS}. Finally, theoretical proposals have
suggested using ultrastrong coupling to enhance the weak nonlinear
terms of the optomechanical Hamiltonian \cite{ClerkLemonde_PRL_2013}
allowing for nonlinear quantum optomechanics.
\begin{acknowledgments}
\emph{G.\,A.\,Peterson acknowledges support from the National Physical
Science Consortium. Official contribution of the National Institute
of Standards and Technology; not subject to copyright in the United
States.}
\end{acknowledgments}

\end{document}


\begin{center} 
{\large \bfseries SUPPLEMENTARY INFORMATION}
\vspace{0.cm}
\end{center}
\begin{center}
\textbf{Ultrastrong parametric coupling between a superconducting cavity\\ and a mechanical resonator}
\end{center}
\begin{center}
{\small
G. A. Peterson,$^{1,2}$ S. Kotler,$^{1,2}$ F. Lecocq,$^{1,2}$ K. Cicak,$^{1}$ X. Y. Jin,$^{1,2}$ \\ R. W. Simmonds,$^{1}$ J. Aumentado,$^{1}$ and J. D. Teufel$^{1}$\\
$^{1}$\emph{National Institute of Standards and Technology, 325 Broadway, Boulder, CO 80305, USA}\\
$^{2}$\emph{Department of Physics, University of Colorado, Boulder, CO 80309, USA}}
\end{center}
\vspace{0.5cm}
\thispagestyle{empty}

\section{Derivation of optomechanical interactions in the ultrastrong coupling
regime }

The Hamiltonian for a cavity optomechanical system is \cite{AspelmeyerKippenbergMarquardt2014}
\begin{equation}
\hat{H}=\hslash\omega_{c}(\hat{x})\left(\hat{n}+\frac{1}{2}\right)+\frac{\hat{p}^{2}}{2m}+\frac{1}{2}m\Omega^{2}\hat{x}^{2},
\end{equation}
where $\omega_{c}$ and $\hat{n}$ are the resonance frequency and
number operator for the microwave cavity mode, and $\hat{p}$, $m$,
$\Omega$, and $\hat{x}$ are the momentum, effective mass, resonance
frequency, and position of the mechanical harmonic oscillator. The
cavity frequency's dependence on mechanical position gives rise to
optomechanical coupling. For small mechanical fluctuations, we use
$\omega_{c}(\hat{x})=\omega_{c}-g_{0}\hat{x}/x_{zp},$ where $g_{0}$
is the single-photon coupling rate and $x_{zp}=\sqrt{\hslash/2m\Omega}$
is the mechanical zero-point fluctuation amplitude, giving rise to
the interaction Hamiltonian in the main text: $\hat{H}_{\text{int}}=\hslash g_{0}\hat{n}\hat{x}/x_{zp}$. 

In the presence of a strong cavity drive at frequency $\omega_{d}=\omega_{c}+\Delta$,
the cavity mode becomes populated with a coherent state of $n_{d}$
driven photons. We expand the Hamiltonian to first order in the fluctuations
$\hat{a}$ around this driven coherent state using $\hat{n}=(\sqrt{n_{d}}e^{i\omega_{d}t}+\hat{a}^{\dagger})(\sqrt{n_{d}}e^{-i\omega_{d}t}+\hat{a})\simeq n_{d}+\sqrt{n_{d}}(\hat{a}^{\dagger}e^{-i\omega_{d}t}+\hat{a}e^{i\omega_{d}t})$.
With $g=g_{0}\sqrt{n_{d}}$ as the parametric coupling, the interaction
Hamiltonian becomes
\begin{equation}
\hat{H}_{\text{int}}=\hslash g_{0}n_{d}\hat{x}/x_{zp}+\hslash g\left(\hat{a}^{\dagger}e^{-i\omega_{d}t}+\hat{a}e^{i\omega_{d}t}\right)\hat{x}/x_{zp}.
\end{equation}
The first term represents a static radiation pressure force that pulls
the capacitor plates closer together. This leads to a power-dependent
shift of the equilibrium position by $2x_{zp}g_{0}n_{d}/\Omega$ and
a shift in the cavity frequency by $-2g^{2}/\Omega$. These shifts
can be absorbed into the definitions of $\hat{x}$ and $\omega_{c}$.
The linearized Heisenberg\textendash Langevin equations of motion
are then given by
\begin{align}
\ddot{\hat{x}}+\Gamma\dot{\hat{x}}+\Omega^{2}\hat{x} & =2\Omega x_{zp}g\left(\hat{a}e^{i\omega_{d}t}+\hat{a}^{\dagger}e^{-i\omega_{d}t}\right)+\frac{\hat{F}_{\text{ext}}}{m},\label{eq:eom_x}\\
\dot{\hat{a}}+i\left(\omega_{c}-\frac{\kappa}{2}\right)\hat{a} & =ig\frac{\hat{x}}{x_{zp}}e^{-i\omega_{d}t}+\sqrt{\kappa}\hat{a}_{\text{in}}.\label{eq:eom_a-1}
\end{align}
where we have included damping rates ($\kappa$ and $\Gamma$) and
external driving ($\hat{a}_{\text{in}}$ and $\hat{F}_{\text{ext}}$)
for each resonator. We can write these equations of motion in compact
matrix form in the Fourier domain as 
\begin{equation}
\mathbf{M}(\omega)\left(\begin{array}{c}
\hat{a}(\omega_{d}+\omega)\\
\hat{x}(\omega)/x_{zp}\\
\hat{a}^{\dagger}(\omega_{d}-\omega)
\end{array}\right)=\left(\begin{array}{c}
i\sqrt{\kappa_{\text{ext}}}\hat{a}_{\text{in}}(\omega_{d}+\omega)\\
\hat{F}_{\text{ext}}(\omega)/2p_{\text{zp}}\\
i\sqrt{\kappa_{\text{ext}}}\hat{a}_{\text{in}}^{\dagger}(\omega_{d}-\omega)
\end{array}\right),
\end{equation}
where $p_{zp}=\sqrt{\hslash m\Omega/2}$ is the zero-point momentum
and the mode-coupling matrix is
\begin{equation}
\mathbf{M}(\omega)=\begin{pmatrix}\chi_{a}^{-1}(\omega_{d}+\omega) & g & 0\\
g & \chi_{x}^{-1}(\omega) & g\\
0 & -g & -\chi_{a}^{-1}(\omega_{d}-\omega)^{*}
\end{pmatrix},\label{eq:opt_M_3node-chi}
\end{equation}
where
\begin{align}
\chi_{a}(\omega) & =1/(\omega-\omega_{c}+i\kappa/2),\\
\chi_{x}(\omega) & =2\Omega/(\omega^{2}-\Omega^{2}+i\omega\Gamma),\label{eq:Mech_susc}
\end{align}
are the complex susceptibility functions for the cavity and mechanical
mode, respectively.

We model the cavity transmission as $T(\omega)=i\sqrt{\kappa_{1}\kappa_{2}}\chi_{a,\text{eff}}(\omega)$,
where $\chi_{a,\text{eff}}\equiv(\mathbf{M}^{-1})_{11}$ is the effective
cavity susceptibility in the presence of optomechanical coupling,
given by
\begin{equation}
\chi_{a,\text{eff}}(\omega)=\left[\chi_{a}^{-1}(\omega)-\frac{g^{2}}{\chi_{x}^{-1}(\omega-\omega_{d})-g^{2}\chi_{a}^{*}(2\omega_{d}-\omega)}\right]^{-1}.\label{eq:chi_a_eff}
\end{equation}
Similarly, the effective mechanical susceptibility $\chi_{x,\text{eff}}\equiv(\mathbf{M}^{-1})_{22}$
is given by 
\begin{equation}
\chi_{x,\text{eff}}(\omega)=\left[\chi_{x}^{-1}(\omega)-g^{2}\chi_{a}(\omega_{d}+\omega)-g^{2}\chi_{a}^{*}(\omega_{d}-\omega)\right]^{-1}.\label{eq:chi_x_eff}
\end{equation}
In the main text, we discuss the normalized effective mechanical susceptibility
$\chi_{m}=\Omega\chi_{x,\text{eff}}/2$, such that $\chi_{m}$ is
dimensionless with a value of $Q_{m}$ on resonance. The mechanical
displacement due to a radiation-pressure force $F$ is then $x(\omega)=\chi_{m}(\omega)F(\omega)/m\Omega^{2}$.
A measurement of the cavity transmission $T(\omega)$ allows us to
extract the mechanical susceptibility from
\begin{equation}
\chi_{m}(\omega-\omega_{d})=\frac{\Omega}{2g^{2}\chi_{a}(\omega)}\left(\frac{T(\omega)}{i\sqrt{\kappa_{1}\kappa_{2}}\chi_{a}(\omega)}-1\right).
\end{equation}

Accounting for multiple mechanical modes is straightforward and relevant
for modeling our data. We assume there are $N$ mechanical modes,
each with a susceptibility $\chi_{i}$, which contains the mode's
resonance frequency and linewidth as in Eq.~(\ref{eq:Mech_susc}),
and a parametric coupling rate $g_{i}$. The extra modes modify the
effective cavity susceptibility (\ref{eq:chi_a_eff}) such that $\chi_{x}$
is replaced by a sum over the mechanical mode susceptibilities, weighted
by their coupling rates: $\chi_{x}\rightarrow\chi_{xN}$, where
\begin{equation}
\chi_{xN}(\omega)=\sum_{i=1}^{N}\left(\frac{g_{i}}{g_{1}}\right)^{2}\chi_{i}(\omega).
\end{equation}
This weighted sum is therefore the effective multimode mechanical
susceptibility to radiation pressure forces. To perform calculations
that account for the higher order modes, we simply replace $\chi_{x}$
in the above discussion with $\chi_{xN}$.

To understand normal-mode splitting and stability, we calculate the
eigenvalues of the optomechanical equations of motion. These are given
by the zeros of the determinant of the mode-coupling matrix: $|\mathbf{M}(\lambda)|=0$.
\begin{equation}
0=\left(\frac{\lambda^{2}-\Omega^{2}+i\Gamma\lambda}{2\Omega}\right)\left(\lambda-\Delta+\frac{i\kappa}{2}\right)\left(\lambda+\Delta+\frac{i\kappa}{2}\right)+2\Delta g^{2}.
\end{equation}
For $\Delta=-\Omega$, the four solutions are
\begin{equation}
\lambda_{\pm\pm}=\frac{\kappa+\Gamma}{4i}\pm\sqrt{\Omega^{2}-\left(\frac{\kappa-\Gamma}{4}\right)^{2}\pm2\Omega\sqrt{g^{2}-\left(\frac{\kappa-\Gamma}{4}\right)^{2}}}.
\end{equation}
Each of the solutions has a corresponding time dependence of $e^{-i\lambda t}$,
so that $\mathrm{Re}[\lambda]$ describes the eigenmode's frequency
and $\mathrm{Im}[\lambda]$ describes its damping rate. If any of
the eigenvalues has a positive imaginary part, the mode amplitude
grows exponentially, and the system is said to be unstable. Using
this criterion, we find that for a red-detuned drive ($\Delta<0$),
and assuming a large mechanical quality factor ($\Gamma\ll\Omega$),
the optomechanical system is unstable for 
\begin{equation}
g^{2}>-\frac{\Omega}{4\Delta}\left(\Delta^{2}+\frac{\kappa^{2}}{4}\right).
\end{equation}

\section{Supplementary data and measurements}

\begin{figure}
\begin{centering}
\includegraphics{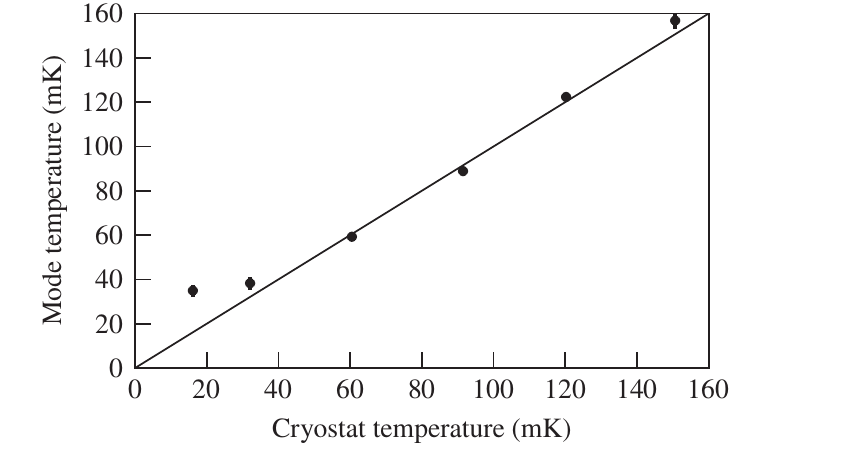}
\par\end{centering}
\caption{Measured mechanical mode temperature as the cryostat temperature is
varied. The mode thermalizes to the cryostat for temperatures above
around $\SI{40}{\milli\kelvin}$. At base temperature, the mechanical
mode reaches $\SI{35}{\milli\kelvin}$. \label{fig:Mechanical-mode-temperature}}
\end{figure}

Figure \ref{fig:Mechanical-mode-temperature} shows the measured mechanical
mode temperature as the cryostat temperature is varied. The data indicates
that the mechanical mode thermalizes with the cryostat down to $\SI{35}{\milli\kelvin}$,
where it becomes thermally decoupled.

\begin{figure}
\begin{centering}
\includegraphics{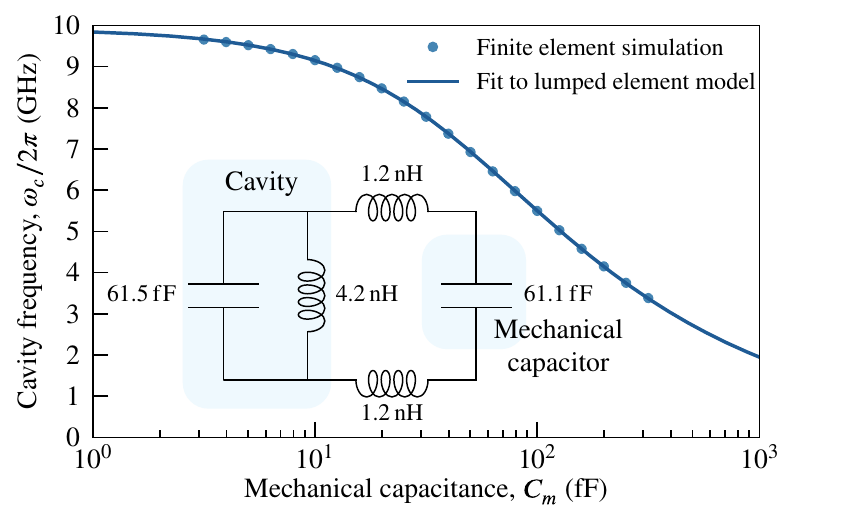}
\par\end{centering}
\caption{Comparison of a three-dimensional finite-element simulation with a
lumped-element circuit model, shown in inset. The simulation solves
for the electromagnetic fields in the cavity as a function of a lumped-element
capacitance, taking into account the sapphire substrate and superconducting
leads connecting the capacitor to the cavity walls. The inset shows
the effective circuit model for our device, where the mechanical capacitance
is $\SI{61.1}{\femto\farad}$, corresponding to a parallel plate separation
of $\SI{29}{\nano\meter}$. \label{fig:comsol}}

\end{figure}

Figure \ref{fig:comsol} shows the result of a finite-element simulation
of the three-dimensional microwave fields in the cavity in the presence
of the sapphire substrate and a lumped element capacitor. The resonant
frequency of the fundamental mode agrees well with a lumped element
circuit model where the cavity mode is modeled as a parallel $LC$
resonance, and the moving capacitor is attached through two inductors,
modeling the thin-film leads in our device.

\begin{figure}
\begin{centering}
\includegraphics{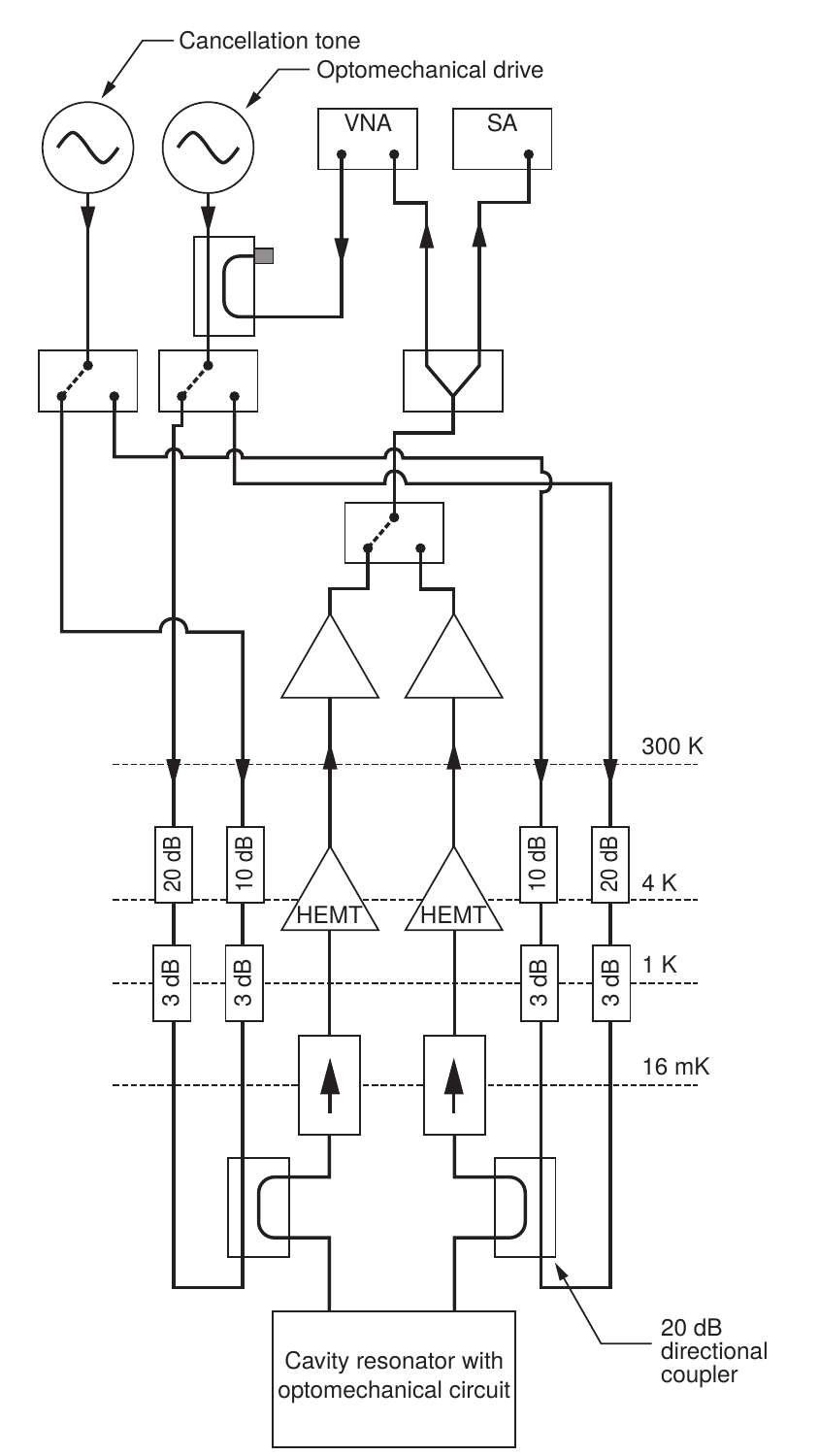}
\par\end{centering}
\caption{Simplified diagram of experimental setup. Two microwave generators
and a vector network analyzer (VNA) act as microwave sources whose
signals are attenuated at low temperature and delivered to the optomechanical
device. The output signals are amplified by cryogenic high electron
mobility transistor (HEMT) amplifiers and room temperature microwave
amplifiers and then measured by the VNA and a spectrum analyzer (SA).
\label{fig:expt_diagram}}
\end{figure}

\begin{table}
\caption{Summary of system parameters. \label{tab:Summary-of-system}}
\begin{centering}
\par\end{centering}
\centering{}\begin{tabular}{ l c S[separate-uncertainty=true,table-parse-only ] s } \toprule  Description & Variable & {Value$/2\pi$} & \multicolumn{1}{c}{Unit}  \\ \midrule Cavity frequency & $\omega_{c}$ & 6.506& \giga\hertz \\ Cavity linewidth & $\kappa$ &  1.2 & \mega\hertz \\ Mechanical frequency & $\Omega$ &  9.696 & \mega\hertz \\ Mechanical linewidth & $\Gamma$ & 31 \pm 1 & \hertz \\ Thermal decoherence rate & $n_{\text{th}}\Gamma$ & 2.4(5) & \kilo\hertz \\ Single-photon coupling rate & $g_{0}$ & 167(2) & \hertz \\ Largest coupling rate & $g$ & 3.83 & \mega\hertz\\ Largest swapping rate & $\Omega_{s}$ & 8.5&\mega\hertz \\ \bottomrule \end{tabular}
\end{table}

See Figure \ref{fig:expt_diagram} for a simplified diagram of the
experimental setup. Table \ref{tab:Summary-of-system} summarizes
the relevant optomechanical parameters of the device.
